\documentclass[twocolumn,showpacs,prl]{revtex4}

\usepackage{graphicx}
\usepackage{dcolumn}
\usepackage{bm}

\begin{document}

\title{Bose Einstein Condensation of incommensurate solid $^4$He}
\author{D. E. Galli and L. Reatto}
\affiliation{
Dipartimento di Fisica, Universit\`a degli Studi di Milano,
Via Celoria 16, 20133 Milano, Italy}
\date{\today}

\begin{abstract}
It is pointed out that simulation computation of energy performed so far cannot be used to decide if
the ground state of solid $^4$He has the number of lattice sites equal to the number
of atoms (commensurate state) or if it is different (incommensurate state).
The best variational wave function, a shadow wave function, gives an incommensurate
state but the equilibrium concentration of vacancies remains to be determined. 
In order to investigate the presence of a supersolid phase
we have computed the one--body density matrix in solid $^4$He
for the incommensurate state by means of the exact Shadow Path Integral Ground State
projector method. We find a vacancy induced Bose Einstein condensation of about 0.23 atoms 
per vacancy at a pressure of 54 bar.
This means that bulk solid $^4$He is supersolid at 
low enough temperature if the exact ground state is incommensurate.
\end{abstract}

\maketitle

Experiments by Kim and Chan\cite{chan,cha2} give evidence for non classical rotational
inertia of solid $^4$He, one hallmark of the supersolid state of matter\cite{legg}.
These results have generated large interest because this would be a novel state characterized,
in a bulk sample, by spontaneous broken translational symmetry
and by a suitable rigidity of the phase of the wave function (wf) giving rise to 
superfluid properties. The standard mechanism for this rigidity is the presence of
Bose Einstein condensation (BEC). Such state with BEC was suggested long ago\cite{andr,ches} as a
possibility for a quantum solid of boson particles. Early theoretical works\cite{legg,andr,ches,sasl}
were based on simplified models so that it was not possible to draw specific predictions
for solid $^4$He. Powerful simulation methods have been applied to study a realistic model of solid
$^4$He in the last two years.
Path Integral Monte Carlo (PIMC) has been applied to study crystalline $^4$He at a finite temperature
and the authors conclude that the superfluid fraction $\rho_s$ at $T=0.2$ K is zero\cite{bern}
and that\cite{cepe} there is no off diagonal long range order (ODLRO), i.e. there
is no BEC, at $T=0.2$ K and 0.5 K. On the other hand the {\it ground state} of crystalline $^4$He
has been studied by variational methods based on shadow wave function (SWF) and this study shows
that ODLRO is present\cite{prbr} for a range of densities above melting with a rather small value of 
BEC fraction $n_0\simeq 0.5 \times 10^{-5}$ at melting density.

One important point to mention is that in these PIMC and SWF computations the crystal is
commensurate in the sense that the number $M$ of lattice sites is equal to the number $N$ of particles,
i.e. $M=N$. 
One finds in the literature\cite{bern,cepe,beam} statements that it is certain that
the ground state of solid $^4$He is commensurate
because microscopic computations\cite{pede,gal5}
have shown that a vacancy increases the energy of the system by at least 15 K and by an even
higher value in the case of an interstitial.
The first purpose of this letter is to present a critical discussion 
of such statements and to examine if the ground state of 
crystalline $^4$He is commensurate (i.e. $N=M$) or if it is incommensurate ($N\not= M$)
in the sense that the lattice parameter inferred from bulk density measurement
differs from the one deduced from Bragg scattering.
Present experiments\cite{simm} do not give evidence for vacancies at low $T$ but new
measurements seem needed to put a stringent bound on ground state vacancies.
The nature of the ground state, commensurate (C) or incommensurate (I),
is a very important point and a phenomenological theory\cite{ande} has shown that the low $T$ 
properties of crystalline $^4$He would be strongly modified should the ground state be incommensurate.
Our conclusion will be that the microscopic computations of solid $^4$He present in literature
do not allow to infer if the ground state is C or I.
Since the presence of vacancies in the ground state cannot be excluded it is important to study
their properties, in particular we study if there is BEC induced by vacancies.
This has been studied variationally\cite{gal1} and here we present an exact computation 
by the projection method SPIGS\cite{gal5} that confirms a vacancy induced BEC.

In the first place we notice that the computations\cite{pede,gal5} at the basis of the estimate of the formation energy 
of a vacancy are actually based on the computation of the {\it ground state energy of two different systems}.
To be specific let us consider the SPIGS computation. The method is based on the application of the
imaginary time evolution operator $\exp\{-\tau\hat{H}\}$ to an assumed trial function,
on a SWF in the case of SPIGS, and on a splitting of this operator 
$\exp\{-\tau\hat{H}\}=[\exp(-{\tau\over P}\hat{H})]^P$
which gives rise to a path integral of {\it linear} polymers. When $\tau$ is large enough a sampling of
the exact ground state of the system is obtained.
Notice that at no stage of the computation, either in the SWF or in the projection procedure,
equilibrium sites of the solid are introduced but the crystalline order, if stable,
arises as spontaneous broken symmetry. With this method two computations are performed, one in which the number $N$
of particles is equal to the number $M$ of lattice sites which fit in the simulation box and satisfy
the periodic boundary conditions (pbc) 
and one in which $N=M-1$. In the second case the simulation shows that the local density continues to have
$M$ maxima with essentially the same degree of crystalline order, as measured by the height of the
Bragg peaks, as in the case $N=M$. 
This means that in the second case the crystalline order is stable with one mobile vacancy and such a state is I.
The SPIGS computation in the two cases gives a converging energy and one example of the evolution
of $E$ as function of $\tau$ is shown in Fig.1 for the fcc lattice.
\begin{figure}
 \includegraphics[angle=0, width=8 cm]{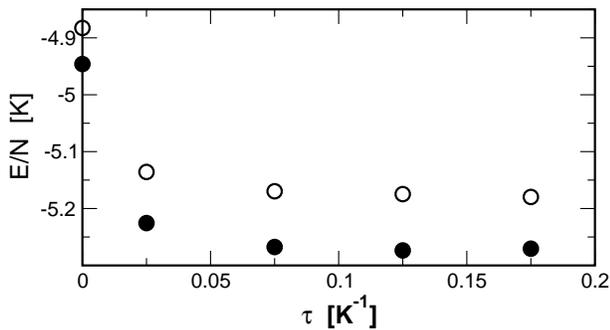}
  \caption{\label{fig1} Energy per particle as function of the evolution in imaginary time $\tau$
of a SPIGS simulation of fcc solid $^4$He in a box with $M=108$.
Filled circles: case C ($N=M$) at $\rho =0.031$\AA$^{-3}$; open circles: case I ($N=M-1$). The largest value of $\tau$
corresponds to 14 projections. Statistical errors are below symbol size.
           }
\end{figure}
Case C is for $N=M=108$ at $\rho =0.031$\AA$^{-3}$
and case I is for $N=M-1=107$.
Starting from a fully optimized SWF\cite{moro} 
just few projections are enough to get convergence in both cases. 
As inter-atomic potential we have used a standard Aziz
potential\cite{aziz}, the time step $\delta=\tau/P$ is (80 K)$^{-1}$
and the pair--product approximation\cite{pimc} has been used
for the imaginary time propagator.
Both wfs are non negative so both 
computations produce ground state energies but the value $E_I$ is slightly larger than $E_C$.
Since we are comparing the energy for two different choices of $N$ it makes no sense to
minimize the energy, both values represent a ground state energy of two periodically repeated small systems.
We conclude that this kind of computation cannot be used to determine if the ground state of bulk solid
$^4$He is C or I unless one is able to extrapolate 
these finite size results to the bulk limit also taking into account the effects
of the pbc. The difference $E_I-E_C$ has been used\cite{gal5} to estimate the formation
energy of an extra vacancy in bulk under the hypothesis of non interacting vacancies
but this is only a derived quantity\cite{comm}.

The fact that the constraints imposed by simulation of a small periodically repeated system
do not allow to determine if the ground state of bulk solid $^4$He is C or I is not surprising. 
In fact also in the classical case it is well recognized\cite{swop} that no direct computation
of the equilibrium concentration $\bar X_v$
of vacancies of a solid at finite temperature is available yet exactly for the same reasons as in the
quantum case: due to finite size of any system that can be simulated and to commensurability effect between
crystal lattice and simulation box the crystal cannot achieve its true equilibrium concentration
of vacancies\cite{swop}.
$\bar X_v$ in the classical case has been obtained only indirectly
by a statistical thermodynamics analysis of an extended system.
In a similar way we can expect to get light on the nature C or I of the ground state of bulk crystalline 
$^4$He only by considering the wf of an extended system,
not of the one which is simulated.

In the framework of variational theory the wfs fall in two categories. One category is a wf which
contains as a factor one-body terms which explicitly break the translational symmetry.
Such wf has great difficulty in describing a vacancy and in fact we are not aware of
any such computation. In addition the nature C or I is built into the wf by construction.
Lastly such wfs with explicitly broken translational symmetry give a worst energy\cite{moro} than the one given by
wfs of the next category. This second category represents translationally invariant wfs
for which the crystalline state arises as spontaneous broken symmetry. 
One such wf is the time honoured Jastrow wf.
The other one is SWF which presently gives the best representation\cite{moro} of the ground state
of solid $^4$He in the sense that it gives the lowest energy.

An important point is that both these translationally invariant wfs give a ground state with a finite
concentration of vacancies and with BEC. In the case of a Jastrow wf this was shown by Chester\cite{ches} and we briefly
repeat here the argument.
Consider a Jastrow wf of a very large system of $\cal N$ particles in volume $\cal V$
\begin{equation}
\Psi_J(R)=\prod_{i<j}^{\cal N}e^{-{1 \over 2} u(r_{ij})}/Q_{\cal N}^{1/2},
\label{uno}
\end{equation}
where $r_{ij}=\mid {\vec r}_i - {\vec r}_j\mid$,
$R=\{{\vec r}_1,..,{\vec r}_{\cal N}\}$
and $Q_{\cal N}$ is the normalization constant
of $\Psi_J^2$, i.e.
\begin{equation}
Q_{\cal N}=\int_{\cal V} dR \prod_{i<j}^{\cal N}e^{-u(r_{ij})}\quad .
\label{due}
\end{equation}
As noticed long ago\cite{mcmi} computation of averages with $\Psi_J^2$ and the normalization $Q_{\cal N}$
have a straightforward interpretation in classical statistical mechanics:
$\Psi_J^2$ coincides with the normalized probability distribution in configurational space
of ${\cal N}$ classical particles at inverse temperature $\beta^*=1/k_BT^*$ and interacting with
a pair potential $v^*(r)$ such that $\beta^*v^*(r)=u(r)$.
$Q_{\cal N}$ is the canonical configurational partition function of this classical system and
its logarithm is proportional to the excess Helmohltz free energy.
It has been proved\cite{reat} that $\Psi_J$ has a finite BEC fraction but it is also known
that the equivalent classical system corresponding to $\Psi_J^2$ is a crystalline solid, when 
the density is large enough, and this solid has a finite concentration of vacancies.
For a classical system the fact that a solid in equilibrium at a finite temperature
has a finite concentration $\bar X_v=({\cal M}-{\cal N})/{\cal N}$ of vacancies,
where ${\cal M}$ is the number of lattice sites, even if a single vacancy has
a finite cost of local free energy derives from the gain in configurational entropy
when the number ${\cal M}-{\cal N}$ of vacancies is proportional to ${\cal N}$\cite{kitt}.
Another way of expressing this is that the configurational partition function $Q_{\cal N}$
of this equivalent classical system has contributions from different pockets in
configurational space, from a pocket $\Omega_0$ in which the positions $\{\vec r_i\}$ of the particles
correspond to vibrations around the equilibrium positions of the commensurate ${\cal N}={\cal M}$ lattice
but also from pockets $\Omega_1$, $\Omega_2$, ... corresponding respectively to ${\cal M}={\cal N}+1$,
i.e. a state with one vacancy, to ${\cal M}={\cal N}+2$ and so on.
It turns out\cite{swop,kitt} that the overwhelming contribution to $Q_{\cal N}$ is associated with
pockets $\Omega_{{\cal M}-{\cal N}}$ with a macroscopic number ${\cal M}-{\cal N}$ of vacancies.
These observations have an immediate interpretation in the quantum case: 
the wf (\ref{uno}) of a bulk system is describing at the same 
time states with no vacancies but also with vacancies and the overwhelming contribution to the 
normalization constant $Q_{\cal N}$ derives from the pockets corresponding to a finite concentration
of vacancies. The simulation of a small system of {\cal N} particles
with pbc is mimicking expectation
values of the quantum Hamiltonian 
of the extended system in a restricted pocket in configurational space, for instance the pocket
$\Omega_0$ of the commensurate state or the pocket $\Omega_1$ of the state with one
vacancy depending if $N=M$ or $N=M-1$.
Notice that in a Monte Carlo (MC) computation the normalization constant (\ref{due})
is never computed explicitly but averages are implicitly normalized to the set of configurations
that are explored in the MC simulation, i.e. to the pocket $\Omega_0$ or $\Omega_1$ that one
has implicitly chosen at the start of the computation by choosing $N=M$ or $N=M-1$.
If we try to estimate the ground state energy per particle $e_G=E_G/{\cal N}$ of a truly
macroscopic system the answer is clear as long as the concentration of vacancies is small
so that they can be considered as independent: If $e_0=E_{M=N}/N$ is the energy per particle
from simulation of the C state and $E_1=Ne_0+\Delta e_v$ the total energy
from simulation of the I state with one vacancy, the inferred ground state energy 
of the extended system is
\begin{equation}
e_G=e_0+\bar X_v \Delta e_v
\label{tre}
\end{equation} 
where $\bar X_v$ is the average concentration of vacancies that should be obtained from an analysis of 
$Q_{\cal N}$ of the extended system. 
This is the true variational energy
of $\Psi_J$ and not $e_0$. Notice that $e_G$ will differ from $e_0$ only by a very small amount
if $\bar X_v$ is well below 1\%.

At present the best variational representation of solid $^4$He is given by a SWF\cite{moro}.
A SWF $\Psi_{SWF}$ contains not only explicit correlations between $\{{\vec r}_i\}$
like in (\ref{uno}) but also indirect correlations built via subsidiary variables,
one for each atom. Also $\Psi_{SWF}^2$ has a classical interpretation, in fact the
normalization of $\Psi_{SWF}^2$ coincides with the configurational partition function of a
classical system of suitable flexible triatomic molecules\cite{swfu}.
For this equivalent classical 
system the concentration $\bar X_v$ of vacancies is finite since in the
previous argument it makes not difference that the ``particles'' are mono-atomic or are molecular
species. We conclude that also $\Psi_{SWF}$ describes a quantum solid
with vacancies in it
and the ground state energy of an extended system is given by eq.\ref{tre}\cite{com3}.

Next question is how big is $\bar X_v$. An estimate of $\bar X_v$ for a Jastrow function has been
performed some years ago\cite{stil} but unfortunately $\Psi_J$ gives an unrealistic
representation of solid $^4$He because $\Psi_J$ gives a much too large localization of atoms.
For $\Psi_{SWF}$ $\bar X_v$ is not known and this is a priority computation for the future.

We consider now the exact ground state as given by SPIGS\cite{gal5}.
The projection maps the quantum problem into an equivalent classical problem of flexible
linear polymers with the number $D$ of monomers equal to $D=2P+1$ where $P$ is the number
of projections. Such classical system has vacancies for any finite $P$ but the
concentration $\bar X_v$ of vacancies might vanish in the limit $P\to \infty$.
Only a study of $\bar X_v$ as function of $P$ will be able to say if ground state vacancies
are present or not in the exact ground state of solid $^4$He.

\begin{figure}
 \includegraphics[angle=0, width=8 cm]{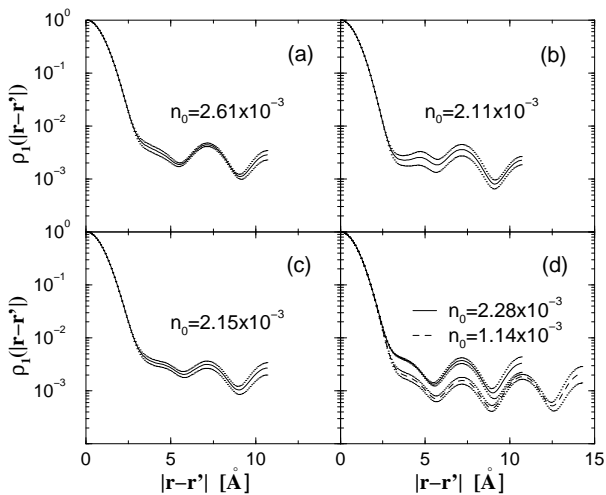}
  \caption{\label{fig3} One-body density matrix $\rho_1$ computed with SPIGS in fcc solid $^4$He at 
$\rho =0.031$\AA$^{-3}$ for different imaginary time evolutions $\tau$:
(a) $\tau=0.025$ K$^{-1}$; (b) $\tau=0.075$ K$^{-1}$; (c) $\tau=0.125$ K$^{-1}$; (d) $\rho_1$ computed with
SWF with $N=M-1=107$ (continuous line) and with $N=M-1=191$ (dashed line).
Dotted lines represent statistical uncertainty.
           }
\end{figure}

Given that vacancies might well be part of the ground state of solid 
$^4$He we present a microscopic calculation of the one--body density matrix $\rho_1$
in presence of vacancies with the exact SPIGS method.
Computation of $\rho_1(\vec r, \vec r')$ is performed by cutting one of the linear polymers at the central 
monomer and sampling the distance $\mid \vec r - \vec r' \mid$
of the two cut ends.
Notice that in SPIGS computations, contrary to the case of PIMC, no exchange moves between
polymers have to be performed and this is a great advantage due to ergodicity problems
arising from exchange moves. In any case
calculation of $\rho_1$ by means of SPIGS in the solid is computationally very intensive
due to the low relaxation (to get converged results one needs more than $10^7$ Monte Carlo steps)
and to the large
number of degrees of freedom (i.e. coordinates of monomers) 
as the imaginary time evolution $\tau$ becomes greater.
There is also the necessity to compute $\rho_1$ as a function of $\tau$ in order to control the convergence.
We have worked with a box which fits $M=108$ fcc lattice sites
and  with $N=107$ so that we have one vacancy out
of 107 atoms which corresponds to $X_v=0.93$\%. In Fig.\ref{fig3} we show the results for three values of
$\tau$\cite{com2} as well as the variational SWF results at density
0.031 \AA$^{-3}$, which 
corresponds to a pressure of about 54 bar.
$\rho_1$ has been computed with $\vec r - \vec r'$ along the nearest neighbor direction ([110] in fcc).
One can see that $\rho_1$ develops a plateau for distances greater than about 5 {\AA}
and this is a signature of BEC. We have estimated
the condensate fraction by averaging the plateau for distance greater than 5.5 {\AA}
and the values are shown in Fig.\ref{fig3}.
The value of $n_0$ is only weakly dependent on $\tau$ and similar to the SWF results.
With SWF it is known that $n_0$ for the hcp lattice is very similar to the fcc one\cite{gal1}.
$\rho_1$ in the plateau region has oscillations and the maxima
correspond to multiples of the nearest neighbor distance.
Since $\rho_1$ is proportional to the probability to destroy a particle at $\vec r$
and to recreate it at $\vec r'$
one can interpret the process giving origin to ODLRO in terms of a sequence of jumps of atoms
which make use of the vacancy. This process is distinct from the vacancy-interstitial pairs
that were found to be important for the commensurate state in a SWF computation\cite{prbr}.
The simulated system is small and with SWF we have computed $\rho_1$ also for a larger system
with $N=M-1=191$ which corresponds to $X_v=0.52$\%.
One sees from Fig.\ref{fig3} the persistence of the oscillations around a finite plateau and it is found
that $n_0$ roughly scale with $X_v$. We are confident that this remains true also for SPIGS
so that we estimate from $n_0$ and the value of $X_v$ of our computation that there is a condensate fraction of
about 0.23 $^4$He atoms per vacancy at the pressure of about 54 bar.
Therefore vacancies are very efficient in inducing BEC.
Using, as an order of magnitude, $T_{BEC}$ of an ideal gas with the effective mass\cite{prlv}
$m^*=0.35\, m_{He}$ we get $T_{BEC}\simeq 11.3\times (\bar X_v)^{2/3}$; for example $T_{BEC}\simeq 0.2$ K
for $X_v=2.3\times 10^{-3}$.
Therefore we expect supersolidity at low $T$ in bulk solid $^4$He if vacancies are present
either as part of the ground state or as non equilibrium effect.

In conclusion we have shown,
on the basis of an exact microscopic theory of solid $^4$He,
the SPIGS projector method,
the presence, at the same time, of spatial order and of BEC
when a finite 
concentration of vacancies is present at $T=0$ K
i.e. if the ground state is incommensurate.
Based on the argument by Leggett\cite{legg} this system would show
non classical rotational inertia effects.
In addition we have shown that the question if the ground state of bulk solid $^4$He is C or I is still
undecided but we noticed that the ground state is I for the best variational wf.
It remains an open problem the quantitative evaluation of the concentration of vacancies
$X_v$ for the SWF and the study of what happens to $X_v$ under projection with the
SPIGS method.

We thank D.M. Ceperley for useful discussions.
This work was supported by the Mathematics Department 
``F. Enriques'' of the Universit\`a degli Studi di Milano and by the INFM Parallel Computing
Initiative.


\begin{thebibliography}{99}
\bibitem{comm} Using the method of ref.12 from $E_B$ and $E_A$ we estimate a vacancy formation energy
at fixed lattice parameter of $8\pm 1$ K which corresponds to $25\pm 2$ K at fixed density.
\bibitem{com2} In the computation of $\rho_1$ we have used a time step $\delta=$ (40 K)$^{-1}$.
The pair product approximation for this $\delta$ is slightly inaccurate but the effect on the energy is only 0.4\%.
\bibitem{com3} At melting the best energy\cite{moro} obtained with a wf with localizing factors is 0.056 K
per atom above the SWF result so that even by allowing for the extra term $\bar X_v \Delta e_v$
the SWF result is below the energy of the wf with localizing factors for any $\bar X_v <$ 0.8\%
where at melting we have estimated $\Delta e_v =6.8\pm 0.9$ K at fixed lattice parameter.
\end{thebibliography}
\end{document}